\documentclass[conference]{IEEEtran}
\IEEEoverridecommandlockouts
\usepackage{cite}
\usepackage{amsmath,amssymb,amsfonts}
\usepackage{algorithmic}
\usepackage{graphicx}
\usepackage{textcomp}
\usepackage{xcolor}
\usepackage{hyperref}
\def\BibTeX{{\rm B\kern-.05em{\sc i\kern-.025em b}\kern-.08em
    T\kern-.1667em\lower.7ex\hbox{E}\kern-.125emX}}
\begin{document}

\title{Quantum Computing for Climate Resilience and Sustainability Challenges
\thanks{Corresponding author: po-heng.lee@imperial.ac.uk}
}

\author{\IEEEauthorblockN{Kin Tung Michael Ho}
\IEEEauthorblockA{\textit{Department of Civil Engineering} \\
\textit{Imperial College London}\\
London, UK \\
kin-tung-michael.ho20@imperial.ac.uk}
\and
\IEEEauthorblockN{Kuan-Cheng Chen}
\IEEEauthorblockA{\textit{Department of Materials} \\
\textit{Imperial College London}\\
London, UK \\
kuan-cheng.chen17@imperial.ac.uk}
\and
\IEEEauthorblockN{Lily Lee}
\IEEEauthorblockA{\textit{Department of Civil Engineering} \\
\textit{Imperial College London}\\
London, UK \\
l.lee23@imperial.ac.uk}
\and
\IEEEauthorblockN{Felix Burt}
\IEEEauthorblockA{\textit{\quad \quad \quad \quad \quad \quad Department of EEE \quad \quad \quad \quad \quad \quad} \\
\textit{Imperial College London}\\
London, UK \\
f.burt23@imperial.ac.uk}
\and
\IEEEauthorblockN{Shang Yu}
\IEEEauthorblockA{\textit{\quad \quad Department of Physics \quad \quad } \\
\textit{Imperial College London}\\
London, UK \\
shang.yu@imperial.ac.uk}
\and
\IEEEauthorblockN{\quad \quad \quad  Po-Heng (Henry) Lee}
\IEEEauthorblockA{\textit{\quad \quad  Department of Civil Engineering } \\
\textit{\quad \quad \quad Imperial College London}\\
\quad \quad \quad London, UK \\
\quad \quad \quad po-heng.lee@imperial.ac.uk}
}

\maketitle

\begin{abstract}
The escalating impacts of climate change and the increasing demand for sustainable development and natural resource management necessitate innovative technological solutions. Quantum computing (QC) has emerged as a promising tool with the potential to revolutionize these critical areas. This review explores the application of quantum machine learning and optimization techniques for climate change prediction and enhancing sustainable development. Traditional computational methods often fall short in handling the scale and complexity of climate models and natural resource management. Quantum advancements, however, offer significant improvements in computational efficiency and problem-solving capabilities. By synthesizing the latest research and developments, this paper highlights how QC and quantum machine learning can optimize multi-infrastructure systems towards climate neutrality. The paper also evaluates the performance of current quantum algorithms and hardware in practical applications and presents realistic cases, i.e., waste-to-energy in anaerobic digestion, disaster prevention in flooding prediction, and new material development for carbon capture. The integration of these quantum technologies promises to drive significant advancements in achieving climate resilience and sustainable development. 
\end{abstract}


\section{Introduction}



The increase of greenhouse gas emissions and subsequent rise in global temperatures has incurred severe environmental and socio-economic consequences. Both mitigation and adaptation strategies are required to achieve net-zero emissions. Consequently, Global Climate Models (GCMs) have become a crucial tool in understanding and predicting the impacts of a changing environment \cite{lupo2013global}. These models simulate the interactions between the atmosphere, oceans, land surface, and ice using mathematical representations to forecast future scenarios based on different greenhouse gas emission pathways.

However, GCMs face substantial challenges that limit their dynamic predictability \cite{lupo2013global}. One critical issue is the need to accurately simulate the complex mosaic of physical, chemical, and biological processes that influence climate over long periods. Many of these processes are either absent or poorly represented in current models, which results in various biases. Additionally, classical computing capacity is constrained at lower resolutions, failing to capture critical regional phenomena such as cloud formation. Using higher-resolution models through downscaling often introduces boundary interactions that propagate errors. The wide range of projected climate responses to increased CO$_2$ highlights significant errors and limitations in these models. Furthermore, GCMs often inadequately account for multiplier effects that can amplify initial climatic impacts, such as those related to solar variations. Despite some improvements from Coupled Model Intercomparison Project 3 (CMIP3) to CMIP5 models, little advancements were made to improve the accuracy of critical climate parameters, highlighting ongoing challenges in climate modeling.

\begin{figure}[!t]
    \centering
    \includegraphics[width=1\linewidth]{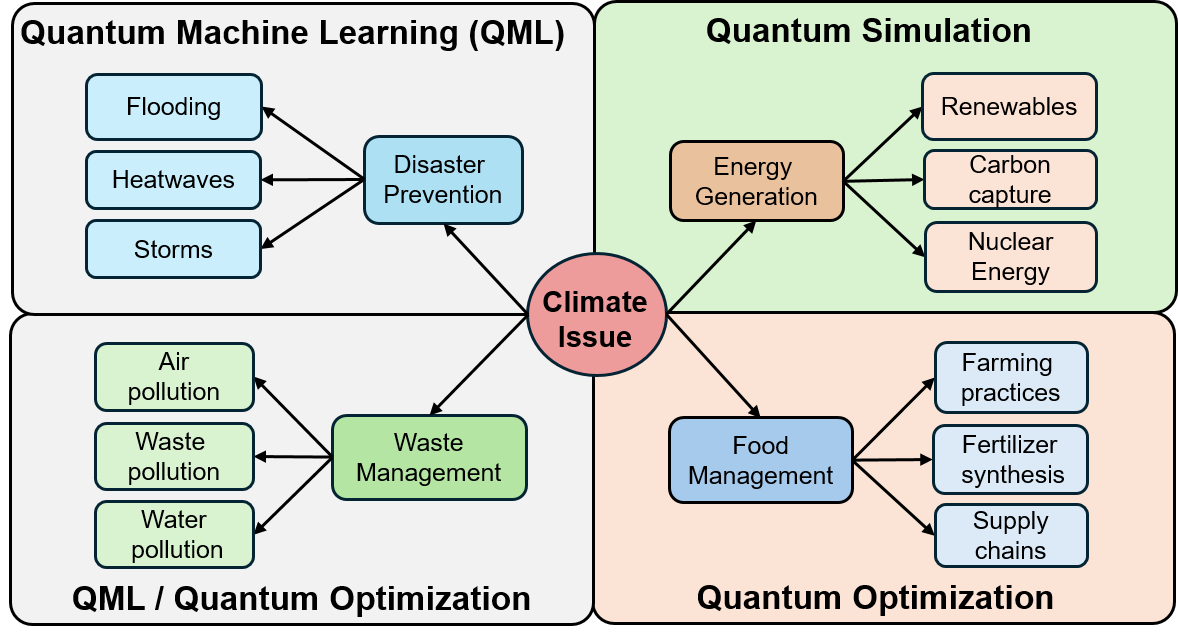}
    \caption{Overview of Quantum Computing applications in climate change mitigation.}
    \label{fig:climate-quantum}
\end{figure}

Machine Learning (ML) and artificial intelligence (AI) are increasingly being applied to various aspects of weather and climate modeling. However, research on the use of ML and AI  using classical computers is still limited compared to its use for nowcasting and weather prediction. This gap is because the available observation period for the earth system is too short to provide enough training data for most applications \cite{de2023machine}. Furthermore, integrating ML into climate models, and more generally natural resource management models, introduces challenges such as addressing instabilities in physical models, increasing the ease of technical integration, and managing memory and computational concerns \cite{de2023machine}. In order to fully realize the impact of ML and AI for enhancing climate predictions and climate-resilience systems, such significant challenges must be addressed alongside the limitations of traditional GCMs. 

Quantum computing (QC), with its enhanced computing power, offers new tools and methodologies to tackle climate challenges more effectively \cite{tennie2023quantum}.  QC is characterized by its potential to solve problems exponentially faster than classical computers\cite{jozsa2003role}. The development of scalable quantum hardware requires the development of quantum algorithms to fully realize the impact of QC for real-world applications \cite{liu2022prospects}. Limited studies have been conducted on utilizing QC to either model climate change and better understand future impacts, or model processes that could improve current natural resource management systems in the face of increasing instability from climate change. 

In this paper, we review the current state-of-the-art in utilizing QC to address challenges in optimization, ML, and simulations, with a particular emphasis on its potential to enhance climate modeling and mitigation strategies for climate change (see Fig. \ref{fig:climate-quantum}). Section 2 delves into the applications of Quantum Optimization in sustainable water treatment and biomass management. Section 3 examines the role of Quantum Machine Learning (QML) in disaster prevention, addressing critical issues such as flooding, heatwaves, and storms. Section 4 investigates the contributions of Quantum Simulation to materials science engineering, focusing on advancements in carbon capture.

\section{Quantum Computing for Optimization}
\subsection{Algorithm Review}
QC holds significant promise for solving complex optimization problems, which are integral to addressing climate change. Among the notable quantum approaches to optimization are quantum annealing (QA) and gate-based quantum algorithms. Quantum annealers, such as those developed by D-Wave Systems, have shown promise in various optimization tasks. It was reported that the Quantum Annealing Continuous Optimization algorithm in \cite{sharabiani2021quantum} was able to optimize biomethane production in anaerobic digestion for the largest biogas producer in Europe, showing advantages over classical computing. 

Quantum annealers are specialized devices designed to solve optimization problems by finding the lowest energy state of a physical system, known as the ground state \cite{rajak2023quantum}. Problems are encoded using the Ising model, where qubits and their interactions form an energy landscape. The annealing process involves gradually reducing quantum fluctuations to guide the system to the ground state, leveraging the adiabatic theorem. This approach allows quantum annealers to tackle complex combinatorial optimization problems more efficiently than classical methods.

However, while QA excels in specific applications, it is less versatile than gate-based quantum algorithms. Gate-based quantum algorithms, particularly the Quantum Approximate Optimization Algorithm (QAOA), offer a more flexible and powerful framework for tackling a broader range of optimization problems\cite{farhi2014quantum}. Unlike quantum annealers, which are specialized for certain tasks, QAOA can be adapted to various problem structures by adjusting its quantum circuit parameters, making it advantageous for complex, multi-objective optimization challenges.

As a hybrid, iterative method, QAOA is designed to solve combinatorial optimization problems. A notable example is the Max-Cut graph problem, which aims to maximize the number of edges between two sets by a partition of the vertices. As an NP-complete graph, reducing the complexity of Max-Cut would help improve the complexity for all problems in NP\cite{farhi2014quantum, zhou2020quantum, xu2024hamiltoniq}. QAOA works by alternating between two types of operations: one that encodes the problem's constraints (the cost Hamiltonian), and one that induces quantum superposition and entanglement (the mixing Hamiltonian). The cost Hamiltonian encodes the objective function of the problem formulated as a ground state corresponds to the optimal solution of the problem. For instance, in the Max-Cut problem, the cost Hamiltonian assigns higher energy to configurations with fewer edges between the two sets, thus incentivizing cuts that maximize the number of edges between them. The mixing Hamiltonian, on the other hand, introduces quantum superposition and entanglement into the system. This enables transitions between different states to prevent the algorithm from getting trapped in a local minimum. By applying these Hamiltonians in alternation, QAOA navigates the solution space more effectively than classical algorithms.





While research into the application of QAOA and other quantum optimizers in water resources management is still limited, there is considerable potential for future research in this domain with substantial impact.  We highlight several areas below. 

\subsubsection{Hydraulic Engineering}
Quantum computers can simulate fluid dynamics using the hydrodynamic Schrödinger equation, which could be used to model three-dimensional turbulent flows in hydraulic engineering to understand and predict fluid dynamics \cite{meng2023quantum}. There are also existing works on using Hamiltonian Simulation to solve the advection equation \cite{brearley_quantum_2024} and various QC methods aimed at approximating Navier-Stokes solutions\cite{ray_towards_2019,gaitan_finding_2020, budinski_quantum_2022}.

\subsubsection{Water Supply Management}
QC can help optimize logistics and supply chain management, which can be applied to water supply management. For example, QAOA can enhance the efficiency of delivery routes, schedule fleet operations, or manage inventory levels more effectively \cite{phillipson2024quantum}. This can lead to more efficient water distribution, minimizing costs and delivery times.

\subsubsection{Wastewater and Waste Treatment Modeling}
Wastewater and waste treatment processes involve numerous parameters that need to be optimized for efficient operation, such as temperature, pH, nutrient concentration, and stirring speed. QAOA can be used to find the optimal combination of these parameters that maximizes the treatment efficiency of the bioreactor.

\subsection{Use Case: Biogas Production in Anaerobic Digestion}
Although QC offers significant advantages on optimizations, there is a notable gap in their application to the fields of water engineering and climate modelling. Limited research has been conducted on the use of either quantum annealers or QAOA for these areas. To the best of our knowledge, \cite{sharabiani2021quantum} is one of the few to investigate the use of QC for optimizing the biomass mix-ratio selection for waste-to-energy in anaerobic digestion. Their work introduces the innovative Quantum Annealing Continuous Optimization (QuAnCO) method to tackle non-discrete optimization problems. QuAnCO utilizes quantum annealing to address the continuous nature of decision spaces in many real-world optimization tasks, such as those related to renewable energy, that traditional QA struggles to handle directly.

The QuAnCO algorithm integrates Trust Region (TR) methods with QA by transforming the TR Newton sub-problem into a Quadratic Unconstrained Binary Optimization (QUBO) problem. This transformation is achieved through three key steps: using a hyper-rectangular shape for the TR, representing each continuous dimension as interval-bounded integers, and binary encoding of these bounded integers. By doing so, QuAnCO allows the application of Ising solvers like D-Wave’s quantum annealer to continuous optimization problems, which are otherwise not directly solvable by such quantum hardware.

The practicality and effectiveness of QuAnCO are demonstrated with a real-world application: optimizing the biomass mix selection for Nature Energy, Europe's largest biogas producer. Biomass mix optimization is a relevant and challenging problem due to the complex, non-convex nature of the cost functions involved. This case study provides strong evidence of the feasibility and performance advantages of using QuAnCO in bio-energy production, showcasing its potential beyond solar energy to broader applications in renewable energy.

Experiments reveal that QuAnCO significantly outperforms traditional optimization methods such as Trust Region Newton (TRN), Conjugate Gradient (CG), and Broyden–Fletcher–Goldfarb–Shanno (BFGS) methods, particularly for complex, non-convex cost functions. For example, QuAnCO-Exact produces a solution that is, on average, approximately 9\% closer to the true minimum (measured from starting point) compared to TRN (p-value: \(4 \times 10^{-5}\)). This performance advantage is most pronounced in the most challenging scenarios studied, indicating that QuAnCO may offer similar benefits for more realistic and complex cost functions, such as the widely used on Anaerobic Digestion Models (ADM1) \cite{batstone2002iwa} for wastewater engineering and biogas production.


\section{Quantum Machine Learning}
\subsection{Algorithm Review}
Quantum Machine Learning (QML) holds significant promise for addressing complex problems with high-dimensional large-scale data in civil engineering, such as predicting the impacts of climate change and managing water resources. Among the various QML algorithms in this domain, the Quantum Support Vector Machine (QSVM)\cite{rebentrost2014quantum} and Quantum-Enhanced Convolutional Neural Networks (QCNN)\cite{cong2019quantum} stand out as potent tools for supervised learning, adaptive decision-making, and pattern recognition tasks, which are essential for predictive modeling and dynamic system management in environmental sciences. Furthermore, Quantum Reinforcement Learning (QRL)\cite{dong2008quantum} and Quantum Train (QT) \cite{liu2024qtrl,lin2024quantum} can be leveraged for complex decision-making processes, offering efficient solutions for large-scale real-world problems through its trial-and-error learning mechanism, which is neither purely supervised nor unsupervised but rather a distinct paradigm of learning based on interactions with the environment.

\subsubsection{Quantum Support Vector Machine }
The QVSM is a quantum algorithm designed to classify data points into one of the categories by finding the optimal hyperplane that separates the classes\cite{li2015experimental,chen2023quantum}. In classical settings, this involves constructing a maximum-margin hyperplane with a normal vector $\mathbf{w}$ that maximizes the distance between the closest points of the two classes, formulated as a quadratic programming problem. The complexity of solving this in the classical setting is polynomial in the size of the data. However, the QSVM leverages QC principles to achieve significant computational advantages. Using quantum techniques such as matrix inversion via the Harrow-Hassidim-Lloyd (HHL) algorithm, QVSM can classify data with a run time that scales logarithmically with the number of features $N$ and training examples $M$\cite{rebentrost2014quantum}.

The core advantage of QSVM is rooted in its ability to perform efficient inner product evaluations and matrix operations. One of the critical equations illustrating this is the dual form of the SVM optimization problem:
\begin{equation}
L(\boldsymbol{\alpha}) = \sum_{j=1}^{M} \alpha_j - \frac{1}{2} \sum_{j,k=1}^{M} \alpha_j K_{jk} \alpha_k,
\end{equation}
where $\alpha_j$ are the Lagrange multipliers and $K_{jk} = \mathbf{x}_j \cdot \mathbf{x}_k$ is the kernel matrix. The quantum version efficiently approximates this matrix inversion, solving for $\boldsymbol{\alpha}$ with the matrix inversion algorithm. This process involves preparing quantum states that represent the training data and exploiting quantum parallelism to perform operations in $\mathcal{O}(\log N \log M)$ time\cite{gentinetta2024complexity}. This results in an exponential speedup over classical methods, particularly beneficial for large-scale datasets common in environmental science applications. 

\begin{figure}[htpb]
    \centering
    \includegraphics[width=1\linewidth]{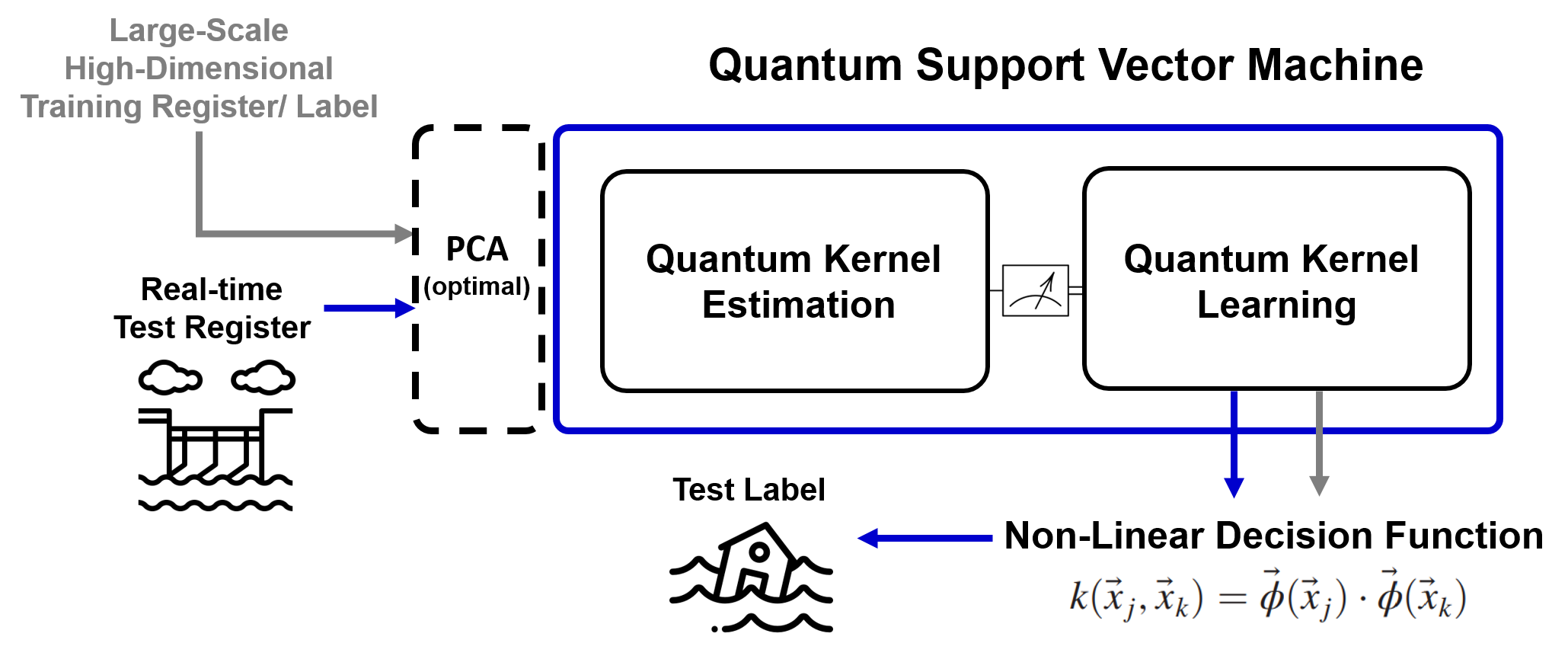}
    \caption{Schematic of a QSVM for flood prediction, illustrating the process from training data through PCA and quantum kernel estimation to final classification.}
    \label{fig:qsvm}
\end{figure}




\subsubsection{Quantum Convolutional Neural Networks}

QCNN are designed to leverage QC for enhancing the capabilities of classical Convolutional Neural Networks (CNNs), widely used in image and pattern recognition tasks\cite{cong2019quantum}. QCNNs apply quantum circuits to perform convolution operations, allowing for the extraction of features from input data with high precision. This enables QCNNs to capture intricate patterns in climate and hydrological data, improving the model's performance in recognizing complex environmental phenomena.

In addition, QCNNs use quantum operations for pooling, a process that reduces the dimensionality of data while preserving essential features. This approach enhances the efficiency and effectiveness of data reduction, facilitating the handling of large-scale environmental datasets. Often, QCNNs combine quantum and classical layers to utilize the strengths of both computing paradigms, processing data more efficiently and accurately. QCNNs' ability to enhance feature extraction and pattern recognition makes them highly valuable for climate change and water engineering applications, such as identifying trends in climate data and predicting environmental changes.

\begin{figure}[htpb]
    \centering
    \includegraphics[width=1\linewidth]{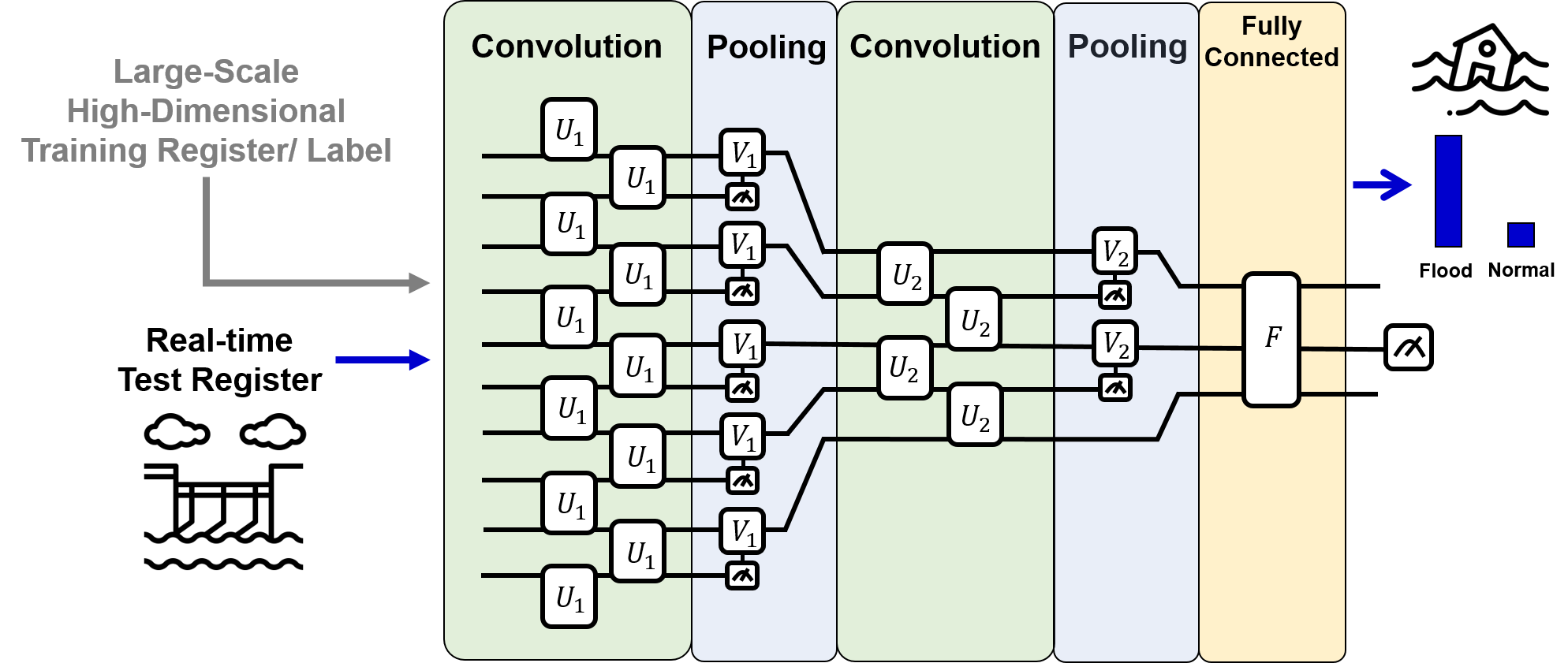}
    \caption{Schematic of a QCNN architecture for flood prediction, illustrating the process from training data through convolution and pooling layers to final flood or normal classification.}
    \label{fig:qcnn}
\end{figure}


\subsection{Use Case: Flood Prediction}
In this section, we utilize a real-world dataset comprising 33 years of time series data from January 1, 1990, to December 31, 2023. This extensive dataset includes measurements from 10 river and dam stations as well as 3 weather stations along the Wupper River in Germany. It is pivotal for developing a predictive model for flood events. In this study, we employ QSVM and QCNN alongside traditional methods such as linear regression and K-Nearest Neighbors (KNN) for benchmarking. Given the large and asymmetric nature of the training data, with only a few instances of flood cases, our benchmarking focuses on sensitivity, particularly the prediction of flood occurrences.

Our benchmarking pipeline incorporates 12 different features related to various reservoirs and accumulated precipitation data, as well as forecast values for 1 day, 2 days, 3 days, 7 days, and 14 days, amounting to a total of 18 features. Principal Component Analysis (PCA) is employed to select the 12 most relevant features for supervised learning tasks. The benchmarking results indicate that KNN and linear regression achieve 0\% sensitivity, highlighting the limitations of traditional supervised learning methods in handling large-scale and high-dimensional unbalanced data. In contrast, QSVM and QCNN achieve sensitivities of 40\% and 80\%, respectively. This demonstrates the potential of quantum-enhanced machine learning for predictions using real-world climate data.

Our study shows that QCNN performs better than QSVM when the number of features (equivalent to the required number of qubits) is small. Furthermore, the validation process for large-scale QSVM can be conducted using the tensor network approach\cite{chen2024cutn}.

\section{Quantum Simulation}
\subsection{Algorithm Review}

One of the most promising near-term noisy intermediate scale quantum (NISQ) algorithms is the variational quantum eigensolver (VQE). VQE is primarily used to simulate and understand physical systems at the quantum level, such as atoms or molecules and their interactions\cite{Bauer_2020}. Current classical computational techniques are inadequate for accurately simulating such systems due to high error rates that scale with the size of the modeled system\cite{Zhang_2022}.   

The general goal of VQE is to calculate the the ground-state energy of a Hamiltonian, which can be used to solve the electronic Schrodinger equation\cite{TILLY2022}. A Hamiltonian is an operator that describes the system’s energy, and the ground-state energy relays the lowest possible energy state of that system. Thus, this calculation is particularly apt for materials science and chemical engineering as the ground state can provide fundamental information regarding the overall energy spectrum of a system and thermodynamic and chemical properties, such as the bonding characteristics of a molecule\cite{ClaryquantumchemistryVQE2023}. As such, the use of VQE could accelerate the development of new materials to both mitigate the effects of climate change and support the transition to a low-carbon economy\cite{FedorovVQEmethod2022}.

VQE is implemented via gate-based quantum circuits comprised of the ansatz, parameters, and a cost function. The purpose of the circuit is to model the physics of a trial wave function using constraints such that the final measured energy approximates the ground-state energy of the Hamiltonian \cite{TILLY2022}. The process begins with initialization of a register of qubits, which is then applied to the ansatz. The ansatz is the structure of ordered gates through which the wave function is encoded into the circuit; it is initially configured with adjustable parameters that are variationally optimized to minimize the ground-state energy. As a result, the ansatz design and initial preparation greatly influences the model accuracy\cite{Sapovacarbonoxidation2022}. The optimization process typically utilizes classical techniques, such as COBYLA or gradient descent, to minimize the cost function. Finally, upon convergence of the cost function, the final measured energy represents the ground-state energy of the Hamiltonian, and the final parameter values represent the parameterized wavefunction used to approximate the ground-state energy \cite{ClaryquantumchemistryVQE2023}. 


\subsection{Use Case: Carbon Capture Materials}
The transition to a low-carbon economy requires extensive research into the use of different materials not only for energy generation and storage, but also for carbon capture. Carbon capture via adsorption to solid sorbents including metal-organic frameworks (MOFs) is a promising method as the energy requirements are low compared to standard post-combustion techniques \cite{stylianou_recent_2015, sumida_carbon_2012}. However, the optimization of these materials requires understanding the specific adsorption mechanisms of CO$_{2}$ with different metals, which represents many-body interactions that are difficult to model with classical computers. As a result, current methods used to predict the properties of new MOFs rely on empirical assumptions of thermodynamic properties that are limited to a small portion of physical scenarios. 

QC could be used to fully simulate the electronic structure of these materials and, in doing so, improve the accuracy and efficiency of the screening of potential MOFs. As such, Greene-Diniz et al. applied VQE to model the interaction of CO$_{2}$ with aluminum nanopores\cite{GDcarbonaluminumVQE2022} and better understand the adsorption energy between CO$_{2}$ and the aluminum cluster (an aluminum atom surrounded by oxygen atoms). This metric that determines how much CO$_{2}$ can be captured and therefore provides information regarding the efficiency of certain materials over others. 

The VQE algorithm utilized the unitary coupled cluster (UCC) ansatz that was applied to a register of qubits representing the electronic orbitals of the adsorbing aluminum atom. The wave function was initially prepared according to a dissociation energy between CO$_{2}$ and the aluminum complex approximated using classical methods. Additionally, different circuits were constructed using 4, 8, and 12 qubits, each modeling a different number of electron orbitals and therefore representing different active spaces with which the CO$_{2}$ could physically adsorb to the aluminum. The results from high-level classical and quantum models were comparable, with predicted bond dissociation energies exhibiting minimal differences (<1 mHa). The authors also found a non-trivial dependence of CO$_{2}$ binding and the non-active sites surrounding the active site, underscoring the versatility of VQE for testing a broad range of physical scenarios that could lead to improved models of this complex binding process in the future. Thus, beyond providing precise output estimations, the application of VQE for materials science could widen the range of physical interactions modeled, which could ultimately expedite the process of understanding and selecting the optimal materials for carbon capture. 

\section{Discussion and conclusion}

The possibilities afforded by QC for climate issues are extensive, yet the amount of research dedicated to these applications is minimal compared to other areas (examples). It is important to start a shift in focus toward both the application of existing QC and QML methods as well as the development of novel QC methods to tackle specific climate-resilience challenges. QuAnCO is a great example of limitations in current methods catalyzing further development. The scope of QuAnCO extends far beyond the original intended use case, as there a various applications of continuous optimization problems that could benefit from transformation to annealing problems, such as the operational parameters in treatment processes, or other, complex, renewable energy sources such as solar and geothermal. In addition, the scope of problems addressable by QAOA is large, particularly concerning route optimization in supply chains for water and food distribution, as well as human transport. 

Similarly, traditional machine learning methods have their limitations, particularly when faced with large volumes of high-dimensional data, such as in kernel transformations for SVM or deep neural networks. It is believed that QML models will be able to express and decode complex patterns at lower computational cost, due inherent complexity of Hilbert space \cite{biamonte_quantum_2017}, potentially allowing us to increase the processing efficiency of high-dimension, high-volume ML models. Climate-resilience problems, such as the flood prediction use case, tend to require such large volumes of data, leading to difficulties for classical ML methods in uncovering the behaviour of systems preceding catastrophe. This highlights the need for QC methods which can perform calculations more efficiently in matrix inversion or pooling operations. We expect the methods used to build flood prediction models could be extended to other disastrous climate events such as severe storms, wildfires and more, as they all occur as a result of complex combinations of various different meteorological and environmental factors, making good targets for QML models.

Quantum simulation methods face issues of their own. Quantum phase estimation (QPE) poses complete solutions to many chemical and material simulation problems, but is out of reach for current quantum computers, hence the reliance on hybrid variational algorithms such as VQE for near-term problems, as variational methods are expected to be more resistant to noise than QPE \cite{TILLY2022}. VQE has issues of its own, however \textemdash VQE is known to work reasonably well when good prior knowledge of the problem is available, most importantly a good initial ansatz \cite{VQEchemicalsystems2022}. If such an ansatz is not available, VQE is unlikely to provide useful results. To optimise algorithms like VQE for climate related problems, this prior knowledge is essential, and will come only from attempted implementation. As QC resources increase, more advanced methods are expected to become feasible. These methods are largely unexplored for climate-related chemical and material problems and hold strong future research prospects. Quantum simulation possibilities also extend beyond material and chemical challenges, showing promise also in improving simulation methods in computational fluid dynamics \cite{brearley_quantum_2024,ray_towards_2019, gaitan_finding_2020,budinski_quantum_2022}. Simulating complex fluid dynamic systems presents a strong path for improving our best weather and climate models. 


In this work, we have provided a brief review on the potential applications of QC in tackling future climate-resilience problems, notably those relating to renewable energy and water management. 


\bibliographystyle{siamurl}
\bibliography{references}

\end{document}